\documentclass[reprint, apl, twocolumn,showpacs,superscriptaddress,groupedaddress]{revtex4}
\usepackage{color,graphicx,calc,url}
\usepackage{dcolumn}
\usepackage{bm}
\usepackage{amssymb}
\usepackage{amsmath}
\usepackage{wasysym}

\begin{document}
\title{Real-time monitoring of the structure of ultra thin Fe$_3$O$_4$ films during growth on Nb-doped SrTiO$_3$(001)}

\author{O.~Kuschel}
\affiliation{Department of Physics and Center of Physics and Chemistry of New Materials, Osnabr{\"u}ck University, 49076 Osnabr{\"u}ck, Germany}
\author{W.~Spiess}
\affiliation{Department of Physics and Center of Physics and Chemistry of New Materials, Osnabr{\"u}ck University, 49076 Osnabr{\"u}ck, Germany}
\author{T.~Schemme}
\affiliation{Department of Physics and Center of Physics and Chemistry of New Materials, Osnabr{\"u}ck University, 49076 Osnabr{\"u}ck, Germany}
\author{J.~Rubio-Zuazo}
\affiliation{SpLine Spanish CRG Beamline at The ESRF, 38043 Grenoble, France}
\author{K.~Kuepper}
\affiliation{Department of Physics and Center of Physics and Chemistry of New Materials, Osnabr{\"u}ck University, 49076 Osnabr{\"u}ck, Germany}
\author{J.~Wollschl{\"a}ger}
\affiliation{Department of Physics and Center of Physics and Chemistry of New Materials, Osnabr{\"u}ck University, 49076 Osnabr{\"u}ck, Germany}
\email{jwollsch@uos.de}

%DESY, Photon Science, Notkestra\ss e 85, D-22607 Hamburg, Germany\\
%$^3$Department of Physics, Bielefeld University, Universit\"atsstra\ss e 25, D-33501, Germany\\
\date{\today}


\begin{abstract}
In this work thin magnetite films were deposited on SrTiO$_3$ via reactive molecular beam epitaxy at different substrate temperatures. 
The growth process was monitored \textit{in-situ} during deposition by means of x-ray diffraction.
While the magnetite film grown at 400$^\circ$\,C shows a fully relaxed vertical lattice constant already in the early growth stages, the film deposited at 270$^\circ$\,C exhibits a strong vertical compressive strain and relaxes towards the bulk value with increasing film thickness. 
Furthermore, a lateral tensile strain was observed under these growth conditions although the inverse behavior is expected due to the lattice mismatch of -7.5\,\%.
Additionally, the occupancy of the A and B sublattices of magnetite with tetrahedral and octahedral sites was investigated showing a lower occupancy of the A sites compared to an ideal inverse spinel structure. 
The occupation of A sites decreases for a higher growth temperature.
Thus, we assume a relocation of the iron ions from tetrahedral sites to octahedral vacancies forming a deficient rock salt lattice.  
\end{abstract}

\maketitle

\begin{figure}[t]
\centering
\includegraphics[]{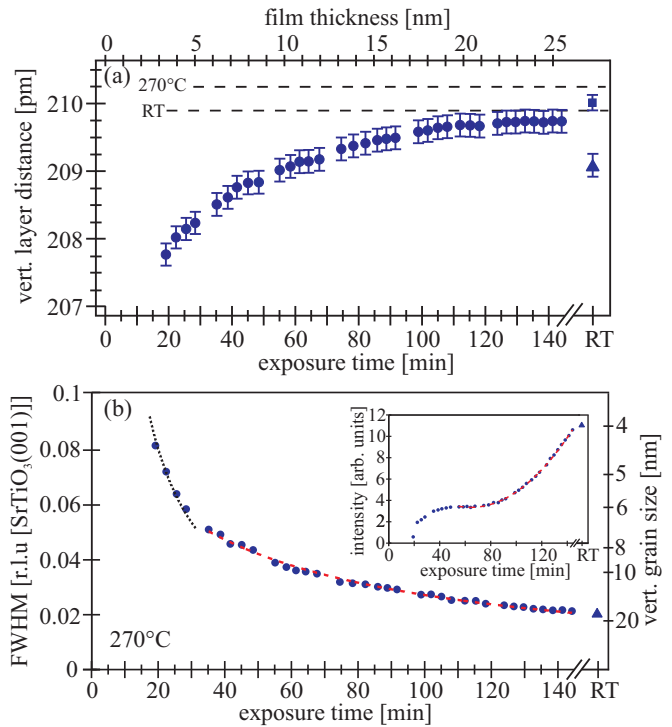}
\caption{(a) Evolution of the vertical layer distance obtained from the position of the Fe$_3$O$_4$(004)$_S$ reflection. The triangular and square symbol mark the at RT measured vertical and lateral lattice distance, respectively. Dashed lines denote the literature values of bulk magnetite at RT and 270$^\circ$\,C.(b) Full width at half maximum of the Fe$_3$O$_4$(004)$_S$ peak as a function of the exposure time. The inset shows the evolution of the peak intensity. The grain size was calculated from the FWHM using the Scherrer formula.}
\label{fwhm270}
\end{figure}

In the rising fields of spintronics \cite{spintronic} and spin caloritronics \cite{spincal} materials with highly spin-polarized carriers are required either for applications based on magnetoresistive effects or on spin-injection \cite{spininjection}. 
For this purpose, the material class of half-metals provides ideal properties with one metallic and another semiconducting or insulating spin channel. 
Here, magnetite (Fe$_3$O$_4$) is one of the intensively studied half-metals \cite{moussy} due to a predicted 100\,\% spin polarization at the Fermi level \cite{halfmet} and a high Curie temperature of 858\,K \cite{BookofIronOxide}, making thin magnetite films, on one hand, particularly suitable for room temperature spintronic applications \cite{mtj, highTMR, spininjector}. 
On the other hand, multilayers of magnetite and platinum enhance the efficiency of thermal generation of spin currents \cite{ramos16} based on the spin Seebeck effect \cite{ramos13,SSE} making Fe$_3$O$_4$ attractive in spin caloritronics as well. \\
Magnetite has a bulk lattice constant of 8.3963\,\AA~\cite{BookofIronOxide} and crystallizes in the inverse spinel structure, where eight tetrahedral (A) sites of the bulk unit cell are only occupied by Fe$^{3+}$ cations while 16 octahedral (B) sites are equally shared by Fe$^{2+}$ and Fe$^{3+}$ cations. 
At about 120\,K bulk magnetite undergoes the so-called Verwey transition, which results in a two-orders-of-magnitude decrease in conductivity and a reduction from cubic to monoclinic crystal symmetry leading to a spontaneous ferroelectric polarization and, thus, multiferroicity \cite{ferroelctric, multiferroic}. 
However, for thin magnetite films this unique transport and magnetic properties as well as structural parameters are strongly influenced by the interaction between the film and the substrate. \\
In this study the influence of the substrate temperature on the growth behavior of thin magnetite films deposited on 0.05\,\% Nb-doped SrTiO$_3$(001) was investigated.
For this system the lattice mismatch between Fe$_3$O$_4$ and SrTiO$_3$ amounts to -7.5\,\%.
Film preparation and characterization were carried out at beamline BM\,25 of the European Synchrotron Radiation Facility (ESRF, Grenoble, France). 
BM\,25 is a bending magnet beamline with a double crystal monochromator consisting of two parallel Si(111) crystals to produce monochromatic beam \cite{BM25}. 
The endstation is equipped with a 2S\,+\,3D diffractometer and an ultra-high vacuum (UHV) chamber. 
The UHV chamber includes thermal evaporation sources, a sample heating device, a LEED (low energy electron diffraction) optics and an x-ray source with a dual Ti/Mg anode and an electrostatic cylinder-sector analyzer to perform x-ray photoelectron spectroscopy (XPS) \cite{BM25_3,BM25_2,BM25_4}. 
The base pressure in the UHV chamber was ~10$^{-10}$\,mbar. 
The set-up design allows to use the sample heating and evaporator and to perform x-ray diffraction (XRD) measurements \textit{during} growth.
For XRD experiments a NaI detector was used.\\
Prior to deposition, the SrTiO$_3$(001) substrates were annealed at 400$^\circ$\,C in 1$\times$10$^{-4}$\,mbar of O$_2$ for 1\,h in order to remove carbon contamination and get well-defined surfaces. 
The crystal surface quality and the chemical cleanness was controlled after each preparation step \textit{in situ} by XPS (Mg K$\alpha_{1/2}$, h$\nu$\,=\,1253.6\,eV) and LEED. 
XPS shows no carbon contamination and LEED reveals quadratic surface symmetry and sharp diffraction spots for the cleaned SrTiO$_3$ substrates. 
Afterwards, thin magnetite films were grown via reactive molecular beam epitaxy (RMBE) (thermal evaporation from pure metal rod in 5$\times$10$^{-6}$~mbar oxygen) at two different substrate temperatures of 270$^\circ$\,C and 400$^\circ$\,C. 
Additionally, for the sample grown at 400$^\circ$\,C the annealing was continued for 30\,min after the evaporation was stopped.  
The resulting film thicknesses were measured by means of x-ray reflectivity (XRR). 
The thickness was determined to be ($25.5\pm0.3$)\,nm and ($10.2\pm0.3$)\,nm for the film grown at 270$^\circ$C and 400$^\circ$C, respectively. 
Hence, the used deposition rate for both samples was ($1.65\pm0.1$)\,\AA/min. 
%The total deposition time for the film grown at 270$^\circ$C was 8700\,sec and the thickness was determined via XRR to be ($25.5\pm0.3$)\,nm. 
%For the at 400$^\circ$\,C grown film the evaporation was stopped after 3600\,sec but the annealing was continued up to 5600\,sec. 
%The resulting film thickness was ($10.2\pm0.3$)\,nm.\\
%\begin{figure}[h]
%\centering
%\includegraphics[]{XPSneu.eps}
%\caption{XP spectra of the Fe2p region of the magnetite thin films grown at 270$^\circ$\,C and 400$^\circ$\,C. For both no charge transfer satellites are visible, indicating the Fe$_3$O$_4$ stoichiometry. Fe2p$_{3/2}$ and Fe2p$_{3/2}$ peaks are located at for magnetite well known binding energies of 710.6\,eV and 723.6\,eV, respectively.}
%\label{XPS}
%\end{figure}
\begin{figure}[t]
\centering
\includegraphics[]{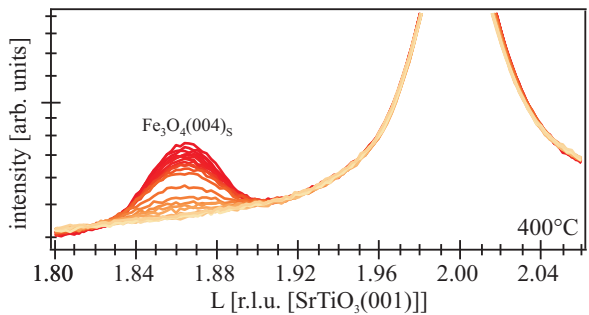}
\caption{XRD measurements along the (00$L$) rod close to SrTiO$_3$(002)$_P$ Bragg reflection. Measurements are performed at a time interval of $\sim$4\,min during deposition at a substrate temperature of 400$^\circ$\,C.} 
\label{XRD}
\end{figure}
Directly after deposition Fe\,2p photoelectron spectra were recorded for both films (not shown here). 
They show no apparent charge transfer satellites, indicating neither an excess of Fe$^{2+}$ nor Fe$^{3+}$ ions \cite{Yamashita,Fuji}. 
Further, the Fe\,2p$_{3/2}$ and Fe\,2p$_{1/2}$ main peaks are located at binding energies of 710.6\,eV and 723.6\,eV corresponding to the well-known values for Fe$_3$O$_4$ \cite{Yamashita}. 
In addition, LEED measurements show a $(\sqrt{2}\times\sqrt{2})$R45$^\circ$ superstructure (not shown here) for both films typical for well ordered magnetite surface \cite{pentcheva, korecki, and97}. 
Combining the results from XPS and LEED we can conclude that both iron oxide films have Fe$_3$O$_4$ stoichiometry and surface structure.\\ 
X-ray diffraction measurements were performed during the deposition of iron oxide at an interval of 3-4\,min. 
Scans along the (00$L$) crystal truncation rod (CTR) were recorded in $\theta-2\theta$ geometry close to the SrTiO$_3$(002)$_P$ and Fe$_3$O$_4$(004)$_S$ Bragg reflections. 
Here, index \textit{P} and \textit{S} denote the indexing for perovskite type (SrTiO$_3$) and spinel type (Fe$_3$O$_4$) bulk unit cells, respectively. 
Since magnetite has almost doubled bulk lattice constant compared to SrTiO$_3$ the Fe$_3$O$_4$(004)$_S$ reflection is located close to the SrTiO$_3$(002)$_P$ Bragg peak but at lower $L$ values.\\
Fig. \ref{XRD} shows the evolution of the Fe$_3$O$_4$(004)$_S$ Bragg peak for the sample grown at 400$^\circ$C. 
In this measurements an intense substrate peak located at $L$\,=\,2 and a much broader Bragg peak at $L\approx$1.86 corresponding to the Fe$_3$O$_4$(004)$_S$ reflection are visible. 
The CTR shows no Laue fringes indicating an inhomogeneous crystalline structure of the film (e.g. inhomogeneous thickness, grains etc.).
With increasing exposure time the intensity of the Fe$_3$O$_4$(004)$_S$ reflection increases while the peak width is decreasing. 
The substrate peak was fitted by a Lorentzian shaped function and the magnetite peak by a Gaussian shaped function to characterize the growth properties. 
Due to low peak intensity, it was only possible to fit the data beyond 15\,min deposition time (equivalent to 2.6\,nm film thickness).
\begin{figure}[t]
\centering
\includegraphics[]{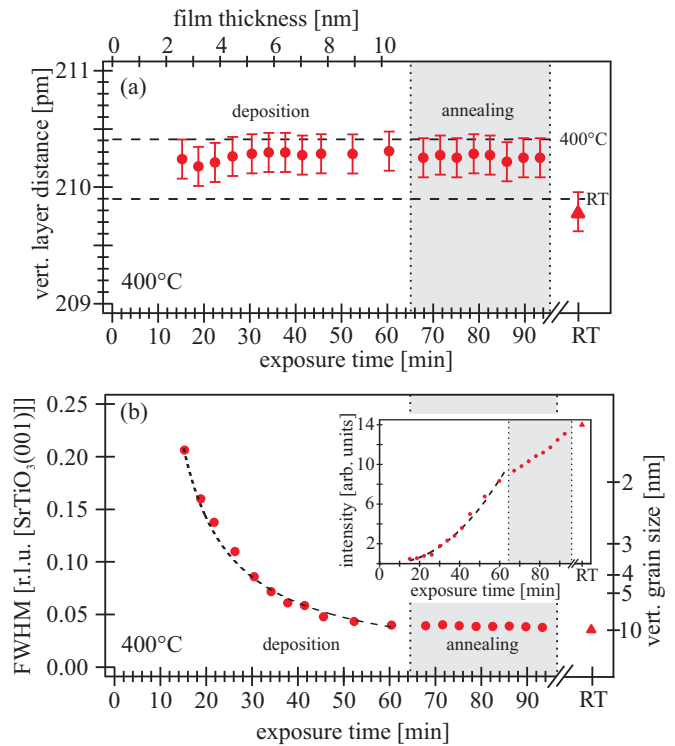}
\caption{(a) Vertical layer distance obtained from the position of the Fe$_3$O$_4$(004)$_S$ reflection. The film thickness was estimated using the evaporation rate. The gray-shaded area denotes the time interval of the subsequent annealing process at 400$^\circ$\,C. The triangular symbol marks the value measured at room temperature (RT). Horizontal dashed lines mark the literature values of bulk magnetite at RT and 400$^\circ$\,C. (b) Full width at half maximum of the magnetite (004)$_S$ peak as a function of the exposure time. The inset shows the evolution of the peak intensity. The grain size was calculated from the FWHM using the Scherrer formula.}
\label{fwhm400}
\end{figure} 
The temporal evolution of the vertical layer distance obtained from the positions of the magnetite diffraction peaks are depicted as a function of the exposure time in Fig. \ref{fwhm400}(a). 
The layer distance remains constant at a value of (210.2\,$\pm\,0.2$)\,pm during the whole deposition and annealing period. 
This value coincides with the value expected for bulk magnetite at 400$^\circ$C taking into account thermal expansion \cite{expcoeff}.
After cooling to room temperature (RT) the resulting layer distance of the magnetite film also coincides within the error tolerance with the bulk value of magnetite \cite{BookofIronOxide}. 
Consequently, the magnetite film deposited at 400$^\circ$\,C grows fully relaxed already at early stages, despite the lattice mismatch between film and substrate of -7.5\%.\\
Fig. \ref{fwhm400}(b) shows the full width at half maximum (FWHM) and the peak intensity (inset) of the Fe$_3$O$_4$(004)$_S$ reflection extracted from curve fitting.
The vertical grain size of the individual steps during the deposition and annealing process was estimated from the FWHM using the Scherrer formula \cite{scherrer}.
Assuming a constant growth rate the time dependence of the FWHM was fitted by a function
\begin{equation}
	FWHM\,=\,\frac{A}{t-t_0}.
	\label{eq1}
\end{equation}
Here, $t_0$ indicates the starting point of ordered growth.
In accordance with the result for the FWHM, the peak intensity follows a parabolic law for $t>t_0$ (cf. inset of Fig.~\ref{fwhm400}(b)). 
%Functions $\frac{A}{t-t_0}$ and $B\cdot(t-t_0)^2$ were fitted to the experimental data of the FWHM and the peak intensity, respectively, recorded during the evaporation process. 
%Here, $t_0$ indicates the starting point of the crystalline growth and the factors $A$ and $B$ depend on the growth rate. 
From the fit of the evolution of the FWHM and intensity of the Fe$_3$O$_4$(004)$_S$ peak an interlayer of 1.0-1.5\,nm thickness was determined.
%Due to a low peak intensity an investigation of the initial growth behavior was not possible.
Here, we assume a high density of point defects and misfit dislocations within this interlayer leading to a fast strain relaxation and, subsequently, the growth of an ordered fully relaxed magnetite film on top.\\
%Assuming the growth rate determined from the grain size and evaporation time during deposition period we receive an amorphous interface layer of 2\,-\,4ML from the evolution of the FWHM and intensity of the magnetite (004)$_S$ peak. 
%Both the FWHM as well as the intensity data are reproduced reasonably well within the applied fitting model.
During the subsequent annealing process ($t>65$\,min) the decrease of the FWHM is negligible while the intensity shows a significant increase pointing to a higher ordering of the magnetite film. 
The resulting increase of the vertical grain size of only 0.5-0.8\,nm is too small compared with the strong increment of the intensity. 
Thus, the strong increase in the intensity during the annealing indicates a lateral ordering of the magnetite film.  
Nevertheless, comparing the vertical grain size calculated by the Scherrer equation and the film thickness obtained from the XRR we estimate a residual distorted interface layer of $\leq$1\,nm. \\
\begin{figure}[t]
\centering
\includegraphics[]{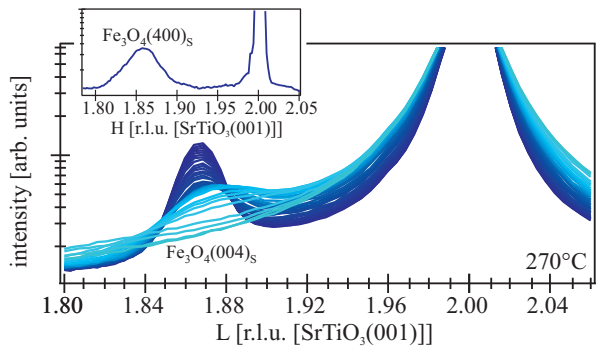}
\caption{XRD measurements along the (00$L$) CTR close to the SrTiO$_3$(002)$_P$ Bragg peak. Measurements are performed at a time interval of $\sim$3\,min during deposition at a substrate temperature of 270$^\circ$\,C. The inset shows the XRD scan along the ($H$00) direction after cooling down to RT.}
\label{XRD2}
\end{figure}
The evolution of the Fe$_3$O$_4$(004)$_S$ Bragg peak for the sample grown at 270$^\circ$C is depicted in Fig.~\ref{XRD2} showing an increase in intensity but a decrease in the peak width with increasing exposure time. 
Here also, no Laue fringes are visible near the Bragg peak pointing to an inhomogeneous crystalline order of the magnetite film.  
In contrast to the film grown at 400$^\circ$C, the Bragg peak shifts to lower $L$ values over the deposition time. 
For detailed analysis the substrate peak was also fitted by a Lorentzian and the Fe$_3$O$_4$ peak by a Gaussian. 
It was not possible to fit the data for the very first 3\,nm film thickness (up to 18\,min) due to negligible peak intensity in the early growth stages. \\  
In Fig. \ref{fwhm270}(a) the vertical lattice constant as a function of the exposure time is presented.
This sample shows a strong strain relaxation behavior towards the bulk value with increasing deposition time.
However, considering the thermal expansion coefficients the bulk value is not completely reached.  
The vertical layer distance increases from (207.8\,$\pm\,0.2$)\,pm after deposition of 3\,nm to a value of (209.7\,$\pm\,0.2$)\,pm at the end of the deposition.  
After cooling down to room temperature the vertical layer distance amounts to (209.1\,$\pm\,0.2$)\,pm, which corresponds to a vertical compressive strain of -0.4\,\%. \\
The lateral lattice constant was determined by measuring the Fe$_3$O$_4$(400)$_S$ Bragg reflection along the ($H$00) direction at room temperature to analyze the structure in more detail (cf. inset of Fig.~\ref{XRD2}). 
The obtained lateral layer distance of (210.2\,$\pm\,0.2$)\,pm exceeds the bulk value of magnetite by 0.14\,\%.
Thus, the Fe$_3$O$_4$ film grown on SrTiO$_3$ at 270$^\circ$C exhibits vertical compressive and lateral tensile strain.
These results are not expected for magnetite on SrTiO$_3$ since the doubled lattice constant of SrTiO$_3$ (3.905\,\AA) is smaller compared to the lattice constant of magnetite (8.3963\,\AA).
Therefore, one expects the inverse behavior, namely lateral compression and vertical tension, due to the lattice mismatch of -7.5\,\%.
The origin of this effect for magnetite deposited at 270$^\circ$C on SrTiO$_3$ is still under discussion.
However, auxetic behavior of this magnetite film, like it was proposed for ultrathin NiFe$_2$O$_4$ films on SrTiO$_3$ \cite{hop15}, can be excluded.\\
The FWHM and the peak intensity (inset) of the Fe$_3$O$_4$(004)$_S$ peak is shown in Fig.~\ref{fwhm270}(b).
Additionally, the vertical grain size estimated from the FWHM using the Scherrer formula \cite{scherrer} is assigned.
For the temporal evolution of the FWHM  Eq.~\ref{eq1} was applied.
The experimental data of the FWHM, however, could only be described by two growth regimes (two different constants A).    
For the first part ($t<30$\,min) of the fit no delay is obtained ($t_0$=0) indicating a continuous reduction of the misfit strain.
The initial fast growth regime is followed by a second stage ($t>30$\,min) where the grains grow more slowly.
In addition, the inset in Fig.~\ref{fwhm270}(b) shows the temporal evolution of the Bragg peak intensity. 
Here, clear conclusions can only be drawn for the second stage of the slow growth. 
The initial constant peak intensity points to the formation of decreasing lateral grain size while we observe a parabolic law for $t>60$\,min. 
The latter agrees well with the observation of growth at 400$^\circ$C and points to a preferential vertical growth of the grains.
The resulting vertical grain size at the end of the deposition amounts to 17\,nm and is $\sim$9\,nm smaller than the film thickness obtained from the XRR measurement.
Probably, these grain boundaries contribute to the relaxation of the strained magnetite film.\\
%Beyond an exposure time of 3300\,sec the measured intensity of the (004)$_S$ peak is well described within the applied model (also der fitfunktion).
%However, up to 3300\,sec the evolution of the intensity exhibits an unexpected behavior showing a stagnation while the FWHM is still decreasing.\\  
In addition, scans along the Fe$_3$O$_4$(22$L$)$_S$ CTR were performed to study separately the occupancy of the A and B sublattices with tetrahedral and octahedral sites, respectively.
\begin{table}
	\centering
		\begin{tabular}{p{1cm}|p{1.4cm}|p{1.4cm}|p{1.4cm}|p{1.4cm}}
\multicolumn{1}{l|}{} & Fe$_3$O$_4$ & 270$^\circ$C   & 400$^\circ$C   & Fe$_{0.75}$O \\  
                      & \footnotesize{(theo.)} & \footnotesize{(exp.)}   & \footnotesize{(exp.)}   & \footnotesize{(theo.)} \\  \hline
$\left.F_{222}\right.$                     & 122.9    & 137.7 & 145.7 & 279.1         \\ \hline
$\left.F_{224}\right.$                     & 135.8    & 122.9 & 116.0 & 0             \\ \hline
$\left.F_{226}\right.$                     & 127.8    & 138.4 & 144.2 & 239.8         \\ \hline
$\varepsilon$                                       & 0        & 0.095 & 0.146 & 1            
\end{tabular}
		\caption{Magnitude of the structure factors for the Bragg peaks of the (22$L$) CTR. Fe$_3$O$_4$ describes an ideal magnetite crystal and Fe$_{0.75}$O an defective rock salt like lattice with the same stoichiometry as magnetite but without tetrahedrally coordinated iron ions. $\varepsilon$ denotes the parameter of disorder calculated following Eq.~\ref{eq2}.}
		\label{epsilon}
\end{table}
%\begin{figure}[t]
%\centering
%\includegraphics[]{epsilon2.eps}
%\caption{Integrated intensity of the Bragg reflection of the Fe$_3$O$_4$(22$L$)$_S$ rod. The inset shows the experimental and calculated intensity ratio between the (224)$_S$ and (222)$_S$ reflection as a function of the occupancy parameter $\varepsilon$.}
%\label{epsilon}
%\end{figure}
%The integrated Bragg peak intensities of the Fe$_3$O$_4$(22$L$)$_S$ CTR are shown in Fig.~\ref{epsilon} for both samples. 
Here, the Fe$_3$O$_4$(224)$_S$ Bragg reflection originates exclusively from the A sublattice with Fe on tetrahedral sites while only Fe cations on octahedral B sites and O anions contribute to the Fe$_3$O$_4$(222)$_S$ and the Fe$_3$O$_4$(226)$_S$ Bragg peaks. 
The latter Bragg peaks were used to determine the Debye-Waller factor so that we could calculate the modulus of the structure factor presented in Tab.~\ref{epsilon}.\\
%To obtain the integrated intensity of each reflection was fitted by a Gaussian function. 
%The intensity of the Bragg peak depends on both the structure factor and the Debye-Waller-factor. 
%We used the Fe$_3$O$_4$ (222)$_S$ and (226)$_S$ reflection to determine the Debye-Waller factor (cf. solid line in Fig.~\ref{epsilon}).\\
%Further, the measured intensities were corrected by a Debye-Waller factor calculated from the intensity ratio of the (222)$_S$ and (226)$_S$ peaks (cf. Fig.~\ref{epsilon}).
Following the model of Bertram et al. the structure factor $F_{HKL}$ of the iron oxide film can be described as a sum of the structure factor of an ideal magnetite $F_{HKL}(Fe_{3}O{_4})$ and of a defective rock salt like structure $F_{HKL}(Fe_{0.75}O)$.
Here, Fe$_{0.75}$O exhibits a Fe$_3$O$_4$ stoichiometry but the vacant B sites of the inverse spinel structure are equally occupied by Fe cations removed from A sites.
Therefore, following the Bragg-Williams theory the structure factor $F_{HKL}$ can be written as 
%Assuming tetrahedral coordinated Fe ions are relocated to octahedral vacancies with a probability $\epsilon$ the structure factor $F(\vec{q})$ can be described as \cite{Berti1} 
\begin{eqnarray}
F_{HKL}=(1-\varepsilon)F_{HKL}(Fe_{3}O{_4})+\varepsilon F_{HKL}(Fe_{0.75}O),
\label{eq2}
\end{eqnarray}
where $\varepsilon$ denotes the parameter of disorder \cite{warren,Berti1}. 
Comparing the experimental values of the Fe$_3$O$_4$(22$L$)$_S$ Bragg peaks with expected values for the ideal structure, the disorder parameter $\varepsilon$ was determined following Eq.~\ref{eq2} (cf. Tab.~\ref{epsilon}).
%We like to emphasize that the disordered rock salt structure has no intensity here.
%In this equation $F_{Fe_{3}O{_4}}(\vec{q})$ is the structure factor of an ideal magnetite and $F_{Fe_{0.75}O}(\vec{q})$ the structure factor of a rock salt like structure with vacant tetrahedral sites but same stoichiometry as Fe$_3$O$_4$. 
%Using kinematic diffraction theory, we can calculate the theoretical values for the probability of occupancy $\varepsilon$ as function of the intensity ratio between (224)$_S$ and (222)$_S$ reflections (cf.~inset of Fig.~\ref{epsilon}). 
%Experimental values for $\varepsilon$ of the two magnetite samples were obtained from intensity ratios obtained from peak fitting.
Both magnetite films show a high but not ideal occupancy of the A sublattice with tetrahedral sites.
The film grown at 270$^\circ$C shows a slightly lower value of $\varepsilon\,=\,0.095$ and, thus, a higher occupancy of the tetrahedral sites, compared to the results reported by Bertram et al. \cite{Berti1} for a well-ordered magnetite film grown at 250$^\circ$C on MgO which could only be obtained for higher growth rates of 3.2\,\AA/s.
%The value of the sample grown at 400$^\circ$C is similar to that reported by Bertram et al. \cite{Berti1} for a well-ordered magnetite film on MgO which could only be obtained for higher growth rates of 3.2\,\AA/s. 
%Despite the strong strain relaxation and lower deposition temperature the sample grown at 270$^\circ$C shows a lower value for $\varepsilon$ compared to the value reported in \cite{Berti1} pointing to a higher occupancy of the tetrahedral sites. \\ 
Despite the fully relaxed growth and a higher deposition temperature the film grown at 400$^\circ$C exhibits a lower ordering of the tetrahedral sublattice ($\varepsilon\,=\,0.146$) compared to magnetite film grown at 270$^\circ$C.\\   
%While the lattice distance of the film deposited at 400$^\circ$C remains constant the distance of the at 250$^\circ$C deposited film relaxes towards the literature value for increasing deposition time.
%\begin{figure}[]
%\centering
%\includegraphics[]{latticeconstant.eps}
%\caption{Evolution of the vertical layer distance of the iron oxide films deposited at a substrate temperature of 270$^\circ$\,C (blue symbols) and 400$^\circ$\,C (red symbols) obtained from the position of the Fe$_3$O$_4$ (004)$_S$ reflection. The vertical doted line denotes the end of evaporation and the start of the annealing for the sample deposited at 400$^\circ$\,C. The triangular symbols mark the at room temperature (RT) measured vertical lattice distance. Doted lines mark the literature values for the layer distance of bulk magnetite at RT, 270$^\circ$\,C and 400$^\circ$\,C \cite{}.}
%\label{latconst}
%\end{figure}
%Due to thermal expansion the lattice constants of both magnetite film and substrate are  Bragg peaks are shifting with increasing
%temperature. Because the thermal expansion coefficient for and iron oxide are almost identical we can use the Bragg reflection as a calibration to compensate the shift caused by thermal expansion.
In summary, the growth process of two magnetite films deposited on SrTiO$_3$(001) at 270$^\circ$C and 400$^\circ$C was monitored by measuring (00$L$) CTRs during deposition. 
The magnetite film grown at 270$^\circ$C exhibits a vertical compressive strain and relaxes continuously over the entire growth process.
Additionally, a lateral tensile strain is obtained excluding auxetic behavior.
Due to a lattice mismatch of -7.5\,\% and, thus, anticipated lateral compressive and vertical tensile strain, this contradicts the behavior expected due to lattice mismatch and requires further investigations.
In contrast, for the sample grown at 400$^\circ$C we assume a strong strain relaxation within the very first few layers followed by a fully relaxed growth regime.
However, magnetite grown at 400$^\circ$\,C shows a lower ordering of the sublattices due to a lower occupancy of the A sites compared to the sample deposited at 270$^\circ$C. 
This points to a relocation of the iron ions from tetrahedral sites to octahedral vacancies forming a deficient rock salt lattice. 
\section*{Acknowledgments}
The authors acknowledge the Deutsche Forschungsgemeinschaft (DFG) via grant no. KU2321/2-1 for financial support. Further, we would like to thank the ESRF for provision of synchrotron radiation in using BM25. %All presented experiments were performed on beamline BM\,25 at the European Synchrotron Radiation Facility (ESRF), Grenoble, France. We are grateful to Juan Rubio-Zuazo and Germán R. Castro at the ESRF for providing assistance in using beamline BM\,25.

%%\begin{thebibliography}{10}
%%\settocbibname{Bibliography}
%%\bibliographystyle{unsrt}
\bibliography{PapervorlageNotes}
%%\end{thebibliography}

\end{document}